\documentclass[12pt]{article}
\usepackage[english]{babel}
\usepackage{amssymb,amsmath}
\usepackage{graphicx}

\usepackage[latin1]{inputenc}

\usepackage{times}
\usepackage{epsf}
\newcommand{\be}{\begin{equation}}
\newcommand{\ee}{\end{equation}}

\vspace{2cm}
\begin{document}
\title{\bf \bf Dispersive Approach to Abelian Axial Anomaly and $\eta-\eta'$ Mixing}
\date{}
\author{Y.~N.~Klopot$^1$\footnote{{\bf e-mail}: klopot@theor.jinr.ru}\;,\,
        A.~G.~Oganesian$^2$\footnote{{\bf e-mail}: armen@itep.ru}\;\, and \
        O.~V.~Teryaev$^1$\footnote{{\bf e-mail}: teryaev@theor.jinr.ru}}
\maketitle
 \hspace{-8mm}
  {$^{1}$\em Joint Institute for Nuclear Research,\;\;Joliot Curie 6, Dubna, Russia, 141980.\\
  $^{2}$Institute of Theoretical and Experimental Physics,\;\;B.Cheremushkinskaya 25, Moscow, Russia, 117218\\
       }

\begin{abstract}
We investigate what can be learnt about the $\eta$ -$\eta'$ mixing by means of  dispersive representation of axial anomaly. 
We show that our method  leads to the strong bounds for the $\eta-\eta'$ mixing angle:
 $\theta = -15.3\ensuremath{^\circ} \pm 1\ensuremath{^\circ}$.
Moreover, our result manifests also a dramatic dependence of the width $\Gamma_{\eta\to 2\gamma}$ on the mixing angle $\theta$.
  This property  explains how the relatively small mixing strongly effects the decay width.

\end{abstract}

\
\section{Introduction}

The problem of the $\eta\eta'$  mixing and particularly the value
of the mixing angle $\theta$ has attracted a lot of interest  for a long period of
time. In the last decade the experimental data  significantly improved (see e.g. \cite{Thomas:2007uy} and references therein).
So the improvement of the accuracy of the theoretical prediction of the mixing angle $\theta$ seems to be important.
Very rich literature is dedicated to this question (see \cite{kp}-\cite{Escribano:2005qq}).

 The well-known estimation, based on Gell-Mann-Okubo mass formulas give the
 value of $\theta$ about -10$^o$ for a quadratic mass formula and
 about  -23$^o$ for a linear case. Obviously, the accuracy of these
 estimations is not high.

In the well-known paper \cite{kp} many years ago by use of the
expansion of the composite operators in the interpolating fields
the values of the $\eta\eta'\to2\gamma$ and
$J/\psi\to\eta(\eta')\gamma$ were considered in the  terms of the octet
currents. The value of the mixing angle $\theta$ was obtained there by
comparison of theoretical results with experimental results (existed
at that time). Two possible mechanisms for the pseudoscalar field
mixing, offered in \cite{zum}, so called mass mixing and
current-mixing models were discussed, and the values of $\theta$
about $-15^o\pm1.8^o$ and $-19.2^o\pm2.2^o$ correspondingly were obtained.

 The analysis of the axial anomaly generated decays
 $\eta(\eta')\to2\gamma$ was also performed in \cite{Donoghue:1986wv} (in the framework of ChPT)
 and \cite{Gilman:1987ax} and the estimate $\theta$ = $-20^o\div 25^o$ was obtained.
 The compehensive  general analysis of the pseudoscalar mesons decays by making use of
 both  abelian and nonabelian anomalies was performed in \cite{Diakonov:1981nv}.
 This approach is based on a standard (differential)
 form of the anomaly and on the pole approximation.

 Another systematic anomaly based approach  was developed in  \cite{Akhoury:1987ed, Ball:1995}, where a large
 set of decay processes was investigated. The mixing angle was
 estimated as $\theta$ = $-17^o \pm 2^o$, which significantly differs from the
 estimation of \cite{Diakonov:1981nv}: $\theta= -9^o$ . The possible explanation of such discrepancy
 is a different role played by the continuum contributions in this two approaches.
The significance of the continuum contribution was explicitly mentioned in  \cite{Ball:1995}. It was noted,
that for $\eta\eta'\to2\gamma$ decay the value $\theta$ = $-17^o \pm 2^o$ can be
obtained only if one takes into account the continuum contribution which cancel the effects of the large
SU(3) breaking  (i.e. the ratio $f_8/f_{\pi}=1.25$), otherwise
the value of $\theta$ grows up to -21$^o$.

The effect of isoscalar pseudoscalar continuum on the mixing angle was also
examined in \cite{Nasrallah:2004ms}. In this approach the masses of the
isoscalar mesons and their decays were discussed. The value of the
mixing angle  was estimated in the range $-30.5^o<\theta~<-18.5^o$.
The value $-17^o<\theta<-13^o$ was  obtained in \cite{Escribano:1999nh}, where phenomenological
 analysis of a set of decays was performed.

All the above results were obtained in  the  mixing scheme
with one angle. The case of the two
mixing angles \cite{Leutwyler:1997yr,Feldmann:1998vh,Kroll:2005sd,Escribano:2005qq}
we will discuss in the section 4.

We can see that all  results are compatible ,
 but with relatively  large uncertainties.

That's why it seems reasonable, and this is the main  aim of this
paper, to use the dispersion form of axial anomaly to find some
model-independent and precise restriction on the value of the mixing
angle. We will develop the method, used in \cite{Ioffe:2007eg},
where a very precise prediction on $\pi^0 \to 2\gamma$ decay was
obtained.

In section 3 we will discuss the one angle octet-singlet mixing scheme
and a very precise restrictions will be obtained.
The scheme with two angles will be discussed in sect. 4. 

\section{$\eta \to 2\gamma$ Decay Width Discrepancy}

%\subsection[Short First Subsection Name]{First Subsection Name}

Consider a matrix element for a transition of the 8th component of the 
%octet 
axial current into two photons with momenta $p$,  $p'$ and polarizations
  $\epsilon_{\alpha}$ , $\epsilon'_{\beta}$:

\be
 T_{\mu \alpha \beta} (p, p') = \langle p,
\epsilon_{\alpha}; p', \epsilon'_{\beta}  \vert J^{(8)}_{\mu 5}
\vert 0 \rangle \;,
\ee

where

\be
 J^{(8)}_{\mu 5} = \frac{1}{\sqrt{6}}(\bar{u} \gamma_{\mu} \gamma_5 u + \bar{d}
\gamma_{\mu} \gamma_5 d - 2\bar{s} \gamma_{\mu} \gamma_5 s)  \;.
\ee

Here $u$, $d$ and $s$ are fields of 
u,d and s 
respective
quarks.\\
We are interested in the decay into two real photons, thus the general form  of\ $T_{\mu \alpha \beta} (p, p')$ can be expressed as:

\be
T_{\mu \alpha \beta}(p, p') = F_1(q^2) q_{\mu}
\epsilon_{\alpha \beta \rho \sigma} p_{\rho} p'_{\sigma}
+ \frac{1}{2} F_2 (q^2) (p_{\alpha} \epsilon_{\mu \beta \rho
\sigma} - p'_{\beta} \epsilon_{\mu \alpha \rho \sigma}) p_{\rho}
p'_{\sigma} \;,
\ee
where $q=p+p'$.

Using  the dispersive approach \cite{Dolgov:1971ri} (see also \cite{Ioffe:2006ww} and references therein)
 one can derive the exact "anomaly sum rule"~\cite{Horejsi:1985qu,Horejsi:1994aj,VeretinTeryaev:1995} :

\be
\label{sumrule}
\int\limits^{\infty}_{0}~ Im~F_1(q^2) dq^2 =
\sqrt{2} \alpha (e^2_u + e^2_d - 2e^2_s) N_c = \sqrt{\frac{2}{3}}\alpha  \;,
\ee
 where $e_u=2/3$,\;   $e_d=e_s=1/3$ are electric charges of $u$, $d$, and $s$ quarks, $N_c=3$ (the number of colors).

Notice, that in QCD this equation does't have any perturbative corrections \cite{Adler:1969er},
and it is expected that it does not have any nonperturbative corrections as well
due to t'Hooft's consistency principle\cite{Horejsi:1994aj}.
It is also important  that at $q^2 \to \infty$ the function $ImF_1(q^2)$ decreases  as $1/q^4$.

Based on the dispersion relation~(\ref{sumrule}) for an axial anomaly,
 a high precision prediction for $\pi^0\rightarrow 2\gamma$ decay was obtained~\cite{Ioffe:2007eg}
 (~1.5\% , mixing $\pi^0$ - $\eta$ was taken into account) and later was confirmed by experiment.
% (Let us note, that in this case taking into account of  pseudoscalar meson mixing provided
% the contribution of few per cent to the well-known naive result.)

Let us first try to saturate the above relation by the $\eta$
contribution only. We use the following standard  definition of the $\eta$
decay constant:
\be
\langle 0 \vert J^{(8)}_{\mu 5} \vert \eta \rangle = i f_{\eta}
q_{\mu} \;.
\ee

The general form of the $\eta$-contribution  to $T_{\mu \alpha
\beta}(p, p')$ can be written as:

\be
 T_{\mu \alpha \beta} (p, p') = - f_{\eta}
\frac{1}{q^2-m^2_{\eta}} \widetilde{A}_{\eta} q_{\mu} \epsilon_{\alpha \beta
\lambda \sigma}p_{\lambda} p'_{\sigma} \;.
\ee
where $\widetilde{A}_{\eta}$ is a constant. So we see, that in the naive approximation,
 when only  $\eta$
contribution is accounted in the l.h.s. of the sum rule for anomaly
relation~(\ref{sumrule}), one can find $\widetilde{A}_{\eta}$

$$
\widetilde{A}_{\eta} = \sqrt{\frac{2}{3}}\frac{\alpha}{\pi}~ \frac{1}{f_{\eta}} \;.
$$

 Then one can easily calculate the decay width $\eta\rightarrow 2\gamma$ \cite{Ioffe:2006ww}:

\be
 \widetilde{\Gamma}_{\eta \to 2\gamma}  = \frac{1}{3}\frac{\alpha^2}{32\pi^3}~
\frac{m^3_{\eta}}{f^2_{\eta}} \;.
\ee

 If we substitute  experimental numbers of $\alpha$, $m_\eta$ and $f_\eta=1.2f_\pi\approx 150 MeV$
 we'll get the numerical value
$$\widetilde{\Gamma}_{\eta \to 2\gamma}  = 0.12\; keV \;,
$$
which is in a striking contradiction  with an experimental value
$$
\Gamma_{\eta \to 2\gamma}  =
0.510\pm0.026\;keV \;.$$

Such a large contradiction
 with exact  anomaly dispersion relation motivates us to consider the 
 effects arising from the other
states contributions to the sum rule. 
%The question is whether
%the effects of mixing can eliminate this contradiction. Moreover, the
 %resolution of this contradiction provides us with strong
%bounds on the mixing parameters. Let us now consider the effects of
%mixing.
We will mainly discuss the mixing of  $\eta$ and $\eta'$ mesons,
because calculation show, that effect of $\pi-\eta$ mixing is found to
be extremely small. We will say a few words about this later.

\section{One Angle $\eta- \eta'$ Mixing Scheme}

Let us now consider the decay $\eta \rightarrow 2\gamma$
 using the dispersive approach and taking into account  mixing of
 $\eta$ and $\eta'$ mesons.
We start with the well-known octet-singlet mixing scheme.

Following \cite{Ioffe:1979rv,Ioffe:1980mx}, let's introduce
nonorthogonal states $\vert P_8 \rangle$ and $\vert P_0 \rangle$ and
the corresponding fields $\varphi_8$, $\varphi_0$, coupled to
  $J^{(8)}_{\mu 5}$ and $J^{(0)}_{\mu 5}$:

$$ \langle 0 \vert J^{(k)}_{\mu 5}\vert P_l \rangle = i
\delta_{kl} f_k q_{\mu}\;, ~~ k = 8,0\,.
$$
%(B.Ioffe, Yad.Fiz. 1979; B.Ioffe, M.Shifman, PL 1980)
Nonorthogonality of the fields $\varphi_8$, $\varphi_0$
corresponds to the non-diagonal term $\Delta H = m^2_{\eta \eta'}
\varphi_8 \varphi_0$ in the effective interaction Hamiltonian. In
the presence of such term the standard PCAC relation is modified
in the following way:

\be \label{PCAC}
\partial_{\mu} J^{(8)}_{\mu 5} = f_{\eta}(m^2_{\eta} \varphi_8 +
m^2_{\eta \eta'} \varphi_0)\;,
\ee

%\partial_{\mu} J^{(0)}_{\mu 5} = f_{\eta'} (m^2_{\eta'} \varphi_0 +
%m^2_{\eta \eta'} \varphi_8).
%$$
The fields $\varphi_8$,$\varphi_0$ are expressed through the physical fields
$\varphi_{\eta}$,~$\varphi_{\eta'}$ as

\begin{gather}
\varphi_8 =  \varphi_{\eta}\cos \theta + \varphi_{\eta'} \sin\theta\;,     \label{phi_mix1}\\
\varphi_0 =  - \varphi_{\eta}  \sin \theta + \varphi_{\eta'}\cos \theta\;. \label{phi_mix2}
\end{gather}

Mixing angle $\theta$ can be expressed from~(\ref{phi_mix1}),(\ref{phi_mix2}) in terms of masses as:
\be \label{mixparam}
\tan {2\theta}=\frac{2m^2_{\eta \eta'}}{m^2_{\eta'} - m^2_\eta}\;.
\ee

Now $Im F_1(q^2)$ is given by the sum of contributions of $\eta$ and  $\eta'$ mesons.
In order to separate the formfactor $F_1(q^2)$, multiply $T_{\mu \alpha \beta} (p, p')$
by $q_\mu /q^2$. Then, taking the imagenary part of $F_1(q^2)$, we get:

$$Im F_1(q^2)=Im~ q_{\mu}\frac{1}{q^2} \langle 2\gamma \mid J^{(8)}_{\mu 5}\mid
0 \rangle$$
$$ = -\frac{f_\eta}{q^2} Im \langle 2\gamma \mid m^2_{\eta} (\varphi_{\eta} \cos
\theta  +\varphi_{\eta'} \sin \theta ) + m^2_{\eta\eta'} (-\varphi_{\eta}\sin \theta
 +  \varphi_{\eta'} \cos \theta)\mid 0 \rangle $$
$$= \pi f_\eta [A_{\eta}\cos \theta\delta (q^2-m^2_{\eta})
 +  A_{\eta'}\frac{m^2_{\eta}}{m^2_{\eta'}} \sin \theta\delta(q^2-m^2_{\eta'})) $$
\be \label{formff1} -A_{\eta}\frac{m^2_{\eta\eta'}}{m^2_{\eta}}\sin\theta\delta(q^2-m^2_{\eta})
 + A_{\eta'}\frac{m^2_{\eta\eta'}}{m^2_{\eta'}} \cos\theta\delta
(q^2-m^2_{\eta'})] \;,
\ee

where
    $A_{\eta}$ ( $A_{\eta'}$) is the amplitude of decay $\eta(\eta')\to 2\gamma$.

    If we employ sum rule (\ref{sumrule}), we obtain:
\be
     \pi f_\eta [A_{\eta}\cos \theta + A_{\eta'}\frac{m^2_{\eta}}{m^2_{\eta'}} \sin \theta
      - A_{\eta}\frac{m^2_{\eta\eta'}}{m^2_{\eta}}\sin\theta
      + A_{\eta'}\frac{m^2_{\eta\eta'}}{m^2_{\eta'}} \cos\theta ]=\sqrt{\frac{2}{3}}\alpha \;.
\ee

Now let's express amplitudes in terms of decay  widths,
employ the relation (\ref{mixparam}) for a mixing parameter $m_{\eta\eta'}^2$ and finally get the equation for
a mixing angle $\theta$:
\be\label{master1}
     \cos \theta + \beta \frac{m^2_{\eta}}{m^2_{\eta'}} \sin \theta
      - \frac{1}{2}(\frac{m^2_{\eta'}}{m^2_{\eta}}-1)\tan2\theta \sin\theta
     +\frac{\beta}{2}(1- \frac{m^2_{\eta}}{m^2_{\eta'}}) \tan2\theta \cos\theta =\xi \;,
\ee
where the dimensionless parameters $\beta$ and $\xi$ were introduced:

\be \label{betaxiparam}
 \beta = \frac{A_{\eta}}{A_{\eta'}}=
\sqrt{\frac{\Gamma_{\eta' \to 2\gamma}}{\Gamma_{\eta \to 2\gamma}}\frac{m_\eta^3}{m_{\eta'}^3}} \;,\;\;
\xi =\sqrt{\frac{\alpha^2 m^3_\eta}{96{\pi}^3\Gamma_{\eta \to 2\gamma}}\frac{1}{f^2_{\eta}}} \;.
\ee

Before starting a numerical analysis, let us discuss the accuracy of the
obtained equation and possible additional contributions.  The other (besides $\eta$ and $\eta'$)
pseudoscalar meson states contributions can arise as some additional terms in the r.h.s of eq.(\ref{formff1}).
But due to the fact, that $ImF_1$
decrease as $1/q^4$ at large $q^2$, from dimensional ground one can easily conclude, that
higher resonances contribution should be suppressed significantly
stronger than $(m_{\eta}/m_{res})^2$, more probably as a fourth power of the mass ratio.
This fact effectively suppresses uncontrolled contributions of higher resonances and continuum
and, consequently,  allow us to reach a higher accuracy.  This is one of the advantages of this method.
The numerical analysis show, that possible influence of this uncontrolled effects on the
solution of the eq.~(\ref{master1}) (i.e. mixing angle $\theta$) is very small (at a few per cent level).

We also calculated the contribution of the $\pi-\eta$
mixing effects in the same way (in fact all  what is necessary for this
was done in  \cite{Ioffe:2007eg}). The pion mixing contribution was found to
be extremely small (less than a per cent, suppressed as $(m_{\pi}/m_{\eta})^4$), that's
why we don't mention it. 
%in  (\ref{master1}) for simplicity.

For a numerical analysis of our anomaly sum rule equation (\ref{master1}) we use the following experimental data (PDG Review 2008) \cite{Amsler:2008zz}:
$m_\eta = 547.853 \pm 0.024$~MeV, $m_{\eta'} = 957.78 \pm 0.24$~MeV,\;
$\Gamma_{\eta \to 2\gamma} = 0.510 \pm 0.026$~keV, \;$\Gamma_{\eta' \to 2\gamma} = 4.30 \pm 0.15$~keV.\;
 Also for a convenience we will present  $f_\eta$   in units of a decay constant of pion  $f_\pi = 130.7$~MeV.
 The only parameter in eq.~(\ref{master1}) that has no a well-defined experimental value, is $f_\eta$.
 If we use a well established prediction of ChPT (see e.g. \cite{Kaiser:1998ds}) $f_\eta=1.28f_\pi$, we  get:

\be
\theta = -15.3\ensuremath{^\circ} \pm 1\ensuremath{^\circ} \;.
\ee

Error bars include  both theoretical and experimental uncertainties which happened of the same order.
It is useful to express $f_\eta$ from eq.~(\ref{master1}) as a function of $\theta$. 
%($f_\eta$ enters parameter $\xi$).
 Remarkably, the function $f_\eta(\theta)$ within the interval of reasonable  values of $f_\eta$ shows
 quite strong dependence on the mixing angle $\theta$ (fig.1). Moreover, expression~(\ref{master1}) manifests
 also a dramatic dependence of the width $\Gamma_{\eta}\to 2\gamma$ on the mixing angle $\theta$.
  This property of the anomaly motivated equations~(\ref{formff1}),~(\ref{master1}) explains how the relatively small mixing
  strongly effects the decay width.

It is interesting also to note, that eq.~(\ref{master1}) as an equation for $\theta$ formally
 has 6 roots for $f_\eta\gtrsim0.5f_\pi$ and only 4 roots for $f_\eta\lesssim0.5f_\pi$ (see fig.2).
 %The other 5 roots (at the reasonable $f_\eta$) are quite distant from our "physical" root:
%  for $f_\eta=1.28f_\pi$ other (besides $-15.3\ensuremath{^\circ}$) roots
% are $-141.0\ensuremath{^\circ}, -75.6\ensuremath{^\circ}, 34.4\ensuremath{^\circ},
%  89.3\ensuremath{^\circ}, 157.2\ensuremath{^\circ}$.
Moreover, the only physical root is among these two disappearing roots.

%  This fact of the existence of other roots is of rather  academical interest,
%    since there are no physical reasons for the mixing angle to be such large.

\section{Two Angle $\eta- \eta'$ Mixing Scheme}

An approach with one mixing angle in study of $\eta- \eta'$ system dominated for decades.
But in the recent years there was a  rise of interest to the mixing scheme with two angles.
It was noticed that taking into account the chiral anomaly through perturbative expansion in ChPT  can lead
to the introduction of two mixing angles in description of the $\eta$- $\eta'$ system \cite{Leutwyler:1997yr}.
There were also performed  some phenomenological analysis of various decay processes in the approach of two
angle mixing scheme \cite{Feldmann:1998vh,Escribano:2005qq}.

In this section we will consider a two angle mixing scheme for the $\eta- \eta'$ system using
the approach developed in the previous section.
We suggest a straightforward generalization of approach  \cite{Ioffe:1979rv,Ioffe:1980mx} by combining PCAC
relation (\ref{PCAC}) with presence of two mixing angles $\theta_1$ and $\theta_2$ (cf. \cite{Feldmann:1998vh} ):

\begin{gather}
\varphi_8 =  \varphi_{\eta}\cos \theta_2 + \varphi_{\eta'} \sin\theta_1\;,\\
 \varphi_0 =  - \varphi_{\eta}  \sin \theta_2 + \varphi_{\eta'} \cos \theta_1\;.
\end{gather}
When mixing angles $\theta_1$ and $\theta_2$ are equal to each other, one comes back to the scheme with one mixing angle.
The anomaly sum rule equation for a two angle mixing scheme now reads:
\be
     \cos \theta_2 + \beta\frac{m^2_{\eta}}{m^2_{\eta'}} \sin \theta_1  - \frac{m^2_{\eta\eta'}}{m^2_{\eta}}\sin\theta_2 +
     \beta\frac{m^2_{\eta\eta'}}{m^2_{\eta'}} \cos\theta_1 = \xi\;,
\ee
where $\beta$ and $\xi$ are defined by (\ref{betaxiparam}) and 
%the mixing parameter is
\be
m^2_{\eta \eta^{'}}=\frac{{m^2_{\eta^{'}}}\sin{2\theta_1} - {m^2_\eta}\sin{2\theta_2}}{2(\cos{\theta_1}\cos{\theta_2} -
\sin{\theta_1}\sin{\theta_2})}\;.
\ee
%
%$$\beta = \frac{A_{\eta}}{A_{\eta'}}=\sqrt{\frac{\Gamma_{\eta' \to 2\gamma}}{\Gamma_{\eta \to 2\gamma}}\frac{m_\eta^3}{m_{\eta'}^3}}\;,
%$$
%
%$$\xi =\sqrt{\frac{\alpha^2 m^3_\eta}{96{\pi}^3\Gamma_{\eta \to 2\gamma}}\frac{1}{f^2_{\eta}}}\;.
%$$

If we use the experimental values for masses $m_\eta$, $m_{\eta'}$
and widths $\Gamma_{\eta \to 2\gamma}$, $\Gamma_{\eta' \to 2\gamma}$ as in the previous section,
 we will get the equation which contains as parameters  angles $\theta_1$, $\theta_2$ and $f_\eta$.
  For physically interesting values of angles we get the plot shown on the fig.3.
 %  Note the weak dependence of angle $\theta_1$ on the angle $\theta_2$ for physical values of $f_\eta$ (e.g. for $f_\eta=1.28f_\pi$ (thick line on fig.3) change of $\theta_2$ from $-30\ensuremath{^\circ}$ to $0\ensuremath{^\circ}$ gives the change of $\theta_1$ only for $6\ensuremath{^\circ}$).
 The resulting plot for a full space of parameters is presented on the fig.4.
We may see that the angles play rather different role. While $\theta_1$ is tightly bounded by the data, like in the case of one 
angle mixing scheme, the sensitivity ro the changes of $\theta_2$ is rather weak.

%***
%
%Moreover, the anomaly is likely to imply the correlation between
%various resonances (and continuum). Such a relations are supported
%by the possibility the write down the analog of the eq() without
%factor $q_{\nu}$.

\section{Conclusions}

It was shown that  the use of dispersive representation of electromagnetic axial anomaly leads
 to tight bounds for the $\eta-\eta'$ mixing angle: $\theta = -15.3\ensuremath{^\circ} \pm 1\ensuremath{^\circ}$.
 This is model independent precise prediction which should be taken into account in future analyses.

One of the main advantages of this method is effective suppression of the higher contribution
 and higher resonances which are usually the main source of uncertainties.

Moreover, our result manifests also a dramatic dependence of the width $\Gamma_{\eta\to 2\gamma}$ on the mixing angle $\theta$.
  This property  explains how the relatively small mixing strongly effects the decay width.

We considered also two angle mixing scheme and and found that rather weakly correlated:
while the first angle is still tightly bounded,
the second one may vary in a rather wide region.

The authors thank D.I. Dyakonov, A.V. Efremov, G. t'Hooft,  B.L. Ioffe and A.V. Kisselev for useful discussions and comments. 
This work was supported in part  by the Russian Foundation of Basic
Research, projects no. 06-02-16905a, 06-02-16215a, 07-02-91557, 08-02-00896a and the funds from EC to the
project "Study of the Strong Interacting Matter" under contract
N0. R113-CT-2004-506078.

\newpage

\begin{figure}
\centerline{\includegraphics[width=0.9\textwidth]{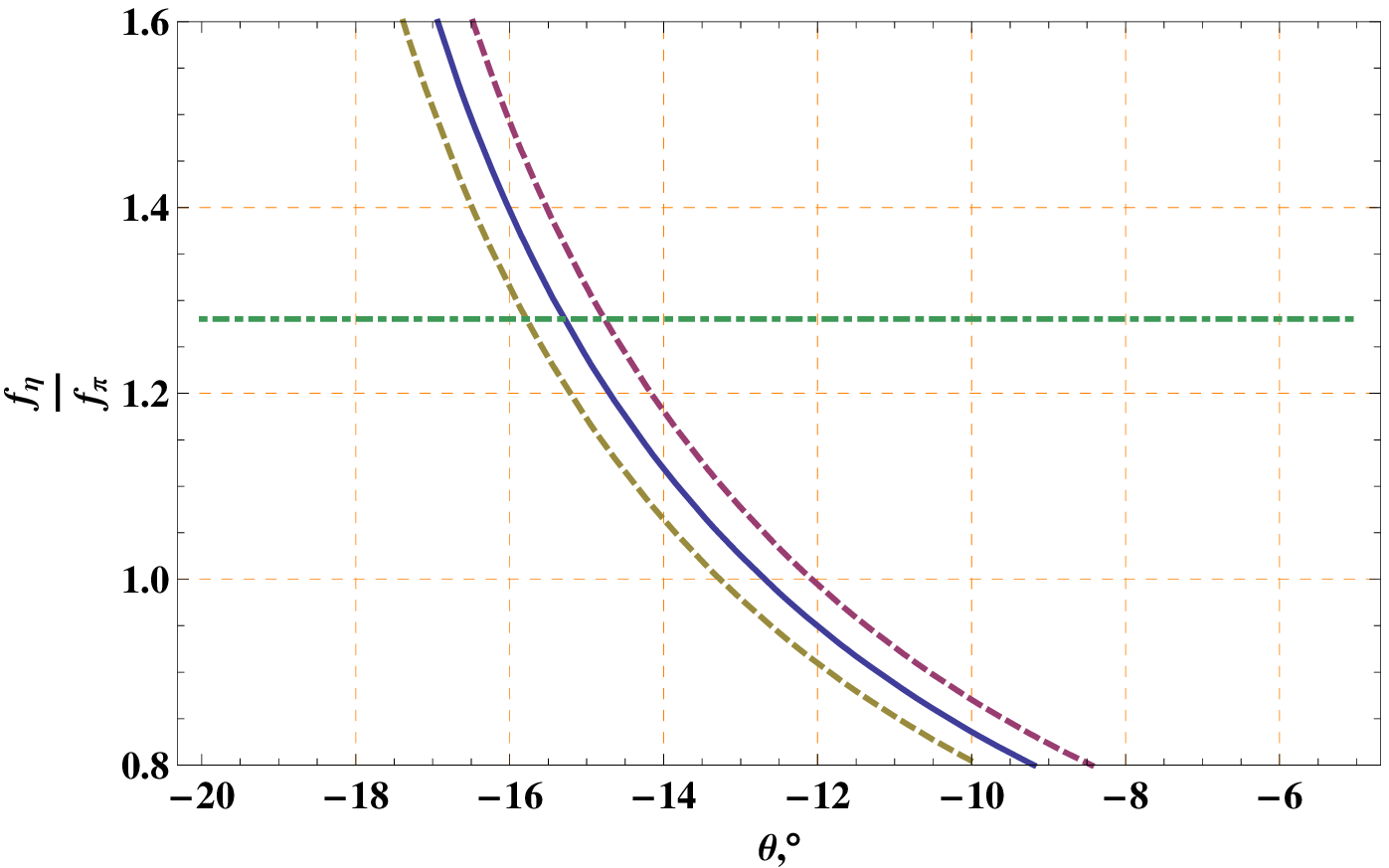}}
\caption{Mixing angle $\theta$ as a function of decay constant $f_\eta$ in the one angle mixing scheme. Dashed lines correspond to the errors of the experimental data input. Dot-dashed horizontal line indicates the  $f_\eta=1.28f_\pi$ level. }
\end{figure}

\begin{figure}
\centerline{\includegraphics[width=0.9\textwidth]{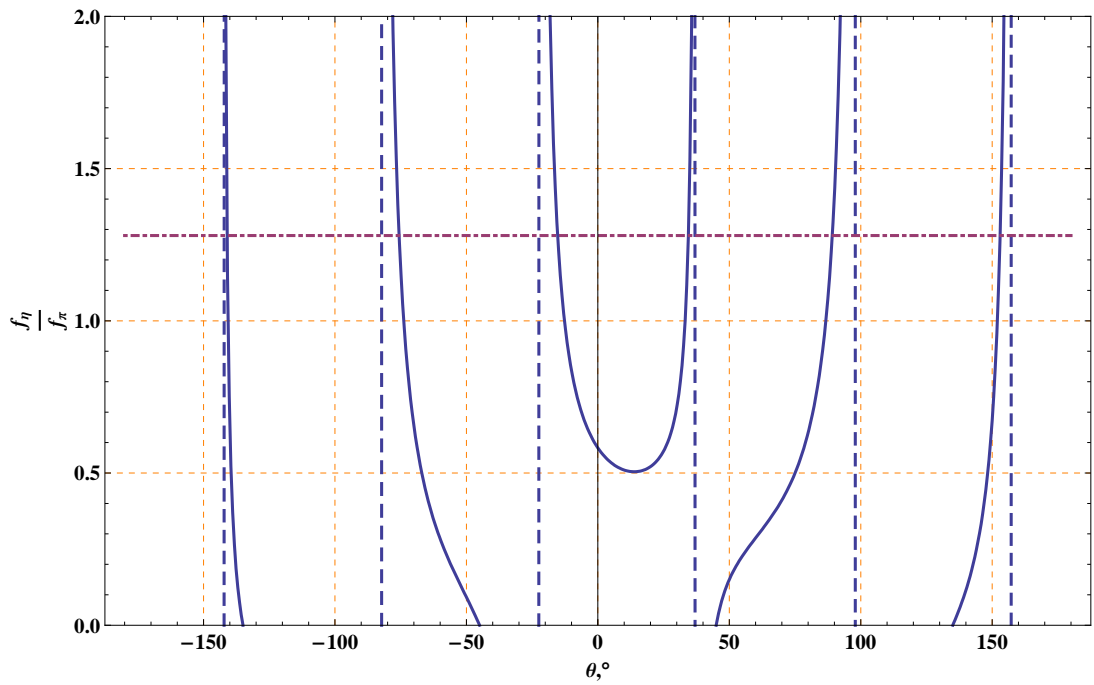}}
\caption{Mixing angle $\theta$ as a function of decay constant $f_\eta$ in the one angle mixing scheme - the full parameter space. Dashed vertical lines indicate asymptotes. Dot-dashed horizontal line indicates the  $f_\eta=1.28f_\pi$ level.}
\end{figure}

\begin{figure}
\centerline{\includegraphics[width=0.75\textwidth]{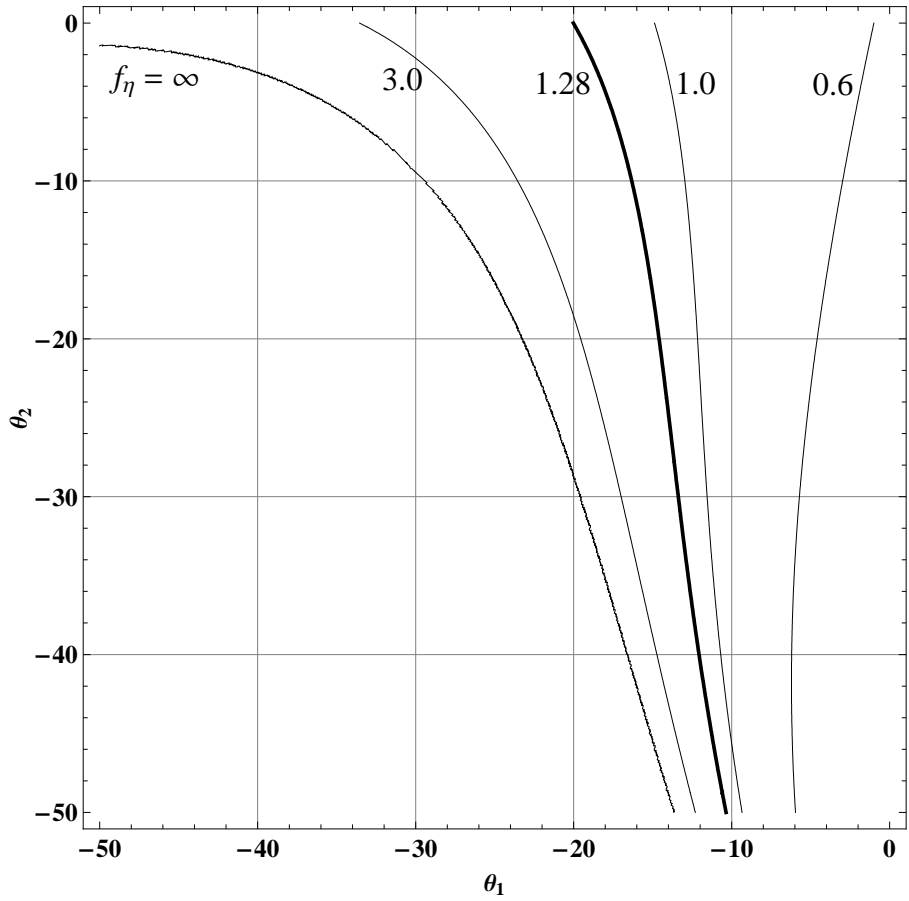}}
\caption{ Mixing angles $\theta_1-\theta_2$ dependence for various $f_\eta$ (as a ratio of $f_\pi$). 
Thick line corresponds to $f_\eta=1.28f_\pi$.  }
\end{figure}

\begin{figure}
\centerline{\includegraphics[width=0.75\textwidth]{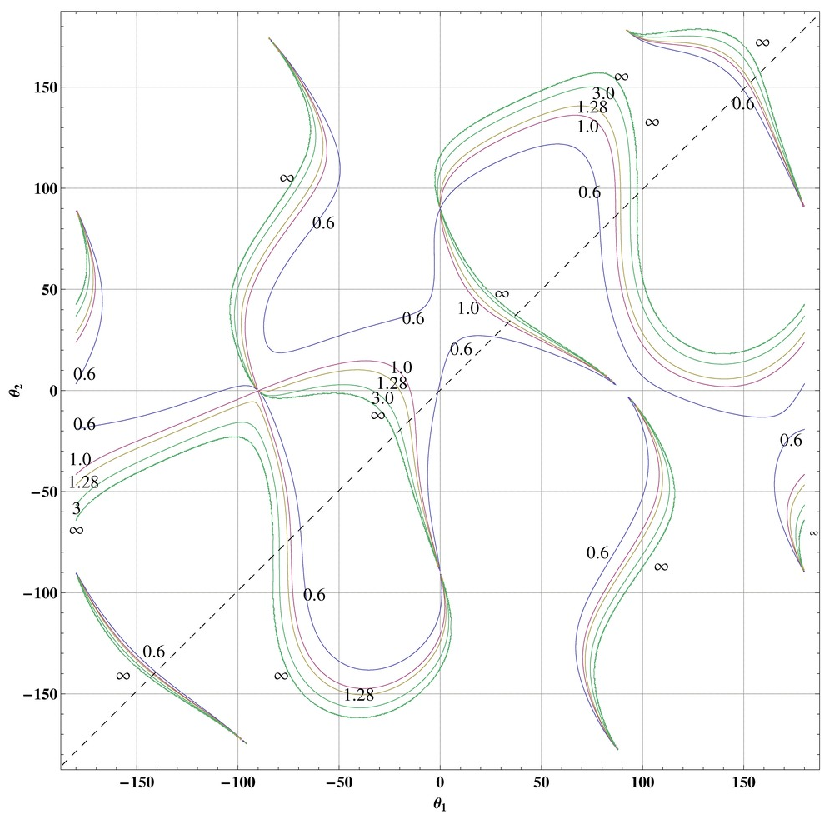}}
\caption{  Two angle mixing scheme - the full parameter space. Different contours correspond to different $f_\eta$ (in units of $f_\pi$, denoted by numbers). The dotted bisector corresponds to the reduction of the one angle mixing scheme ($\theta_1=\theta_2$) }
\end{figure}
%\\

  % Keep the summary *very short*.


\begin{thebibliography}{99}

%\cite{Thomas:2007uy}
\bibitem{Thomas:2007uy}
  C.~E.~Thomas,
  %``Composition of the Pseudoscalar Eta and Eta' Mesons,''
  JHEP {\bf 0710}, 026 (2007)
  [arXiv:0705.1500 [hep-ph]].
  %%CITATION = JHEPA,0710,026;%%

\bibitem{kp}
A.V.~Kisselev and V.A.~Petrov,
Z.\ Phys.\ C{\bf 58}, 595-600 (1993)

\bibitem{zum}
 N.~M.~Kroll, T.~D.~Lee and B.~Zumino,
  %``Neutral Vector Mesons and the Hadronic Electromagnetic Current,''
  Phys.\ Rev.\  {\bf 157}, 1376 (1967).

%\cite{Donoghue:1986wv}
\bibitem{Donoghue:1986wv}
  J.~F.~Donoghue, B.~R.~Holstein and Y.~C.~R.~Lin,
  %``Chiral Loops In Pi0, Eta0 $\to$ Gamma Gamma And Eta Eta-Prime Mixing,''
  Phys.\ Rev.\ Lett.\  {\bf 55}, 2766 (1985)
  [Erratum-ibid.\  {\bf 61}, 1527 (1988)].
  %%CITATION = PRLTA,55,2766;%%

  %\cite{Gilman:1987ax}
\bibitem{Gilman:1987ax}
  F.~J.~Gilman and R.~Kauffman,
  %``The Eta Eta-Prime Mixing Angle,''
  Phys.\ Rev.\  D {\bf 36}, 2761 (1987)
  [Erratum-ibid.\  D {\bf 37}, 3348 (1988)].
  %%CITATION = PHRVA,D36,2761;%%



\bibitem{Akhoury:1987ed}
  R.~Akhoury and J.~M.~Frere,
  %``eta, eta-prime MIXING AND ANOMALIES,''
  Phys.\ Lett.\  B {\bf 220}, 258 (1989).
  %%CITATION = PHLTA,B220,258;%%

\bibitem{Ball:1995}
  P.~Ball, J.~M.~Frere and M.~Tytgat,
  %``Phenomenological evidence for the gluon content of eta and eta-prime,''
  Phys.\ Lett.\  B {\bf 365}, 367 (1996)
  [arXiv:hep-ph/9508359].
  %%CITATION = PHLTA,B365,367;%%
  %\cite{Akhoury:1987ed}

%\cite{Diakonov:1981nv}
\bibitem{Diakonov:1981nv}
  D.~Diakonov and M.~I.~Eides,
  %``Massless Ghost Pole In Chromodynamics And The Solution Of The U(1)
  %Problem,''
  Sov.\ Phys.\ JETP {\bf 54}, 232 (1981)
  [Zh.\ Eksp.\ Teor.\ Fiz.\  {\bf 81}, 434 (1981)].
  %%CITATION = ZETFA,81,434;%%


%\cite{Pham:1990tv}
\bibitem{Pham:1990tv}
  T.~N.~Pham,
  %``TWO PHOTON DECAY OF THE ETA MESON IN BROKEN SU(3) x SU(3),''
  Phys.\ Lett.\  B {\bf 246}, 175 (1990).
  %%CITATION = PHLTA,B246,175;%%

%\cite{Escribano:1999nh}
\bibitem{Escribano:1999nh}
  R.~Escribano and J.~M.~Frere,
  %``Phenomenological evidence for the energy dependence of the eta eta'  mixing
  %angle,''
  Phys.\ Lett.\  B {\bf 459}, 288 (1999)
  [arXiv:hep-ph/9901405].
  %%CITATION = PHLTA,B459,288;%%

%\cite{Nasrallah:2004ms}
\bibitem{Nasrallah:2004ms}
  N.~F.~Nasrallah,
  %``Glue content and mixing angle of the eta - eta' system: The effect of  the
  %isoscalar 0- continuum,''
  Phys.\ Rev.\  D {\bf 70}, 116001 (2004)
  [Erratum-ibid.\  D {\bf 72}, 019903 (2005)]
  [arXiv:hep-ph/0410240].
  %%CITATION = PHRVA,D70,116001;%%


%\cite{Feldmann:1998vh}
\bibitem{Feldmann:1998vh}
  T.~Feldmann, P.~Kroll and B.~Stech,
  %``Mixing and decay constants of pseudoscalar mesons,''
  Phys.\ Rev.\  D {\bf 58}, 114006 (1998)
  [arXiv:hep-ph/9802409].
  %%CITATION = PHRVA,D58,114006;%%

%\cite{Kroll:2005sd}
\bibitem{Kroll:2005sd}
  P.~Kroll,
  %``Isospin symmetry breaking through pi0-eta-eta' mixing,''
  Mod.\ Phys.\ Lett.\  A {\bf 20}, 2667 (2005)
  [arXiv:hep-ph/0509031].
  %%CITATION = MPLAE,A20,2667;%%

%\cite{Escribano:2005qq}
\bibitem{Escribano:2005qq}
  R.~Escribano and J.~M.~Frere,
  %``Study of the eta eta' system in the two mixing angle scheme,''
  JHEP {\bf 0506}, 029 (2005)
  [arXiv:hep-ph/0501072].
  %%CITATION = JHEPA,0506,029;%%

%\cite{Kaiser:1998ds}
\bibitem{Kaiser:1998ds}
  R.~Kaiser and H.~Leutwyler,
  %``Pseudoscalar decay constants at large N(c),''
  arXiv:hep-ph/9806336.
  %%CITATION = HEP-PH/9806336;%%

%\cite{Leutwyler:1997yr}
\bibitem{Leutwyler:1997yr}
  H.~Leutwyler,
  %``On the 1/N-expansion in chiral perturbation theory,''
  Nucl.\ Phys.\ Proc.\ Suppl.\  {\bf 64}, 223 (1998)
  [arXiv:hep-ph/9709408].
  %%CITATION = NUPHZ,64,223;%%

%\cite{Dolgov:1971ri}
\bibitem{Dolgov:1971ri}
  A.~D.~Dolgov and V.~I.~Zakharov,
  %``On Conservation of the axial current in massless electrodynamics,''
  Nucl.\ Phys.\  B {\bf 27}, 525 (1971).
  %%CITATION = NUPHA,B27,525;%%

%\cite{Horejsi:1985qu}
\bibitem{Horejsi:1985qu}
  J.~Horejsi,
  %``On Dispersive Derivation Of Triangle Anomaly,''
  Phys.\ Rev.\  D {\bf 32}, 1029 (1985).
  %%CITATION = PHRVA,D32,1029;%%

%\cite{Horejsi:1994aj}
\bibitem{Horejsi:1994aj}
  J.~Horejsi and O.~Teryaev,
  %``Dispersive Approach To The Axial Anomaly, The T'hooft's Principle And QCD
  %Sum Rules,''
  Z.\ Phys.\  C {\bf 65}, 691 (1995).
  %%CITATION = ZEPYA,C65,691;%%


%\cite{Veretin:1995}
\bibitem{VeretinTeryaev:1995}
  O.~L.~Veretin and O.~V.~Teryaev,
   Yad.\ Fiz. {\bf 58}, 2266 (1995)


%\cite{Ioffe:2006ww}
\bibitem{Ioffe:2006ww}
  B.~L.~Ioffe,
  %``Axial anomaly: The modern status,''
  Int.\ J.\ Mod.\ Phys.\  A {\bf 21}, 6249 (2006)
  [arXiv:hep-ph/0611026].
  %%CITATION = IMPAE,A21,6249;%%

%\cite{Adler:1969er}
\bibitem{Adler:1969er}
  S.~L.~Adler and W.~A.~Bardeen,
  %``Absence of higher order corrections in the anomalous axial vector
  %divergence equation,''
  Phys.\ Rev.\  {\bf 182}, 1517 (1969).
  %%CITATION = PHRVA,182,1517;%%

%\cite{Ioffe:2007eg}
\bibitem{Ioffe:2007eg}
  B.~L.~Ioffe and A.~G.~Oganesian,
  %``Axial anomaly and the precise value of the pi0 --> 2gamma decay width,''
  Phys.\ Lett.\  B {\bf 647}, 389 (2007)
  [arXiv:hep-ph/0701077].
  %%CITATION = PHLTA,B647,389;%%

%\cite{Ioffe:1979rv}
\bibitem{Ioffe:1979rv}
  B.~L.~Ioffe,
  %``Masses Of Light Quarks And Interaction Of Low-Energy Eta Mesons. (In
  %Russian),''
  Yad.\ Fiz.\  {\bf 29}, 1611 (1979).
  %%CITATION = YAFIA,29,1611;%%

%\cite{Ioffe:1980mx}
\bibitem{Ioffe:1980mx}
  B.~L.~Ioffe and M.~A.~Shifman,
  %``Decays Psi-Prime $\to$ J / Psi Pi0 (Eta) And Quark Masses,''
  Phys.\ Lett.\  B {\bf 95}, 99 (1980).
  %%CITATION = PHLTA,B95,99;%%

%\cite{Amsler:2008zz}
\bibitem{Amsler:2008zz}
  C.~Amsler {\it et al.}  [Particle Data Group],
  %``Review of particle physics,''
  Phys.\ Lett.\  B {\bf 667}, 1 (2008).
  %%CITATION = PHLTA,B667,1;%%







\end{thebibliography}
\end{document}